\documentclass[%
 reprint,
nofootinbib,
 amsmath,amssymb,
 aps,
prl
]{revtex4-1}

\usepackage{todonotes}
\usepackage{graphicx}
\usepackage{dcolumn}
\usepackage{bm}
\usepackage{caption}
\usepackage{subcaption}

\begin{document}

\preprint{APS/123-QED}

\title{Arc-Like Variable Bunch Compressors}
\thanks{This work was partially funded by STFC grant ST/P002056/1 ``The Cockcroft Institute of Accelerator Science and Technology".}%

\author{Peter H. Williams}
 \email{peter.williams@stfc.ac.uk}
  \affiliation{STFC Daresbury Laboratory \& Cockcroft Institute, Warrington WA4 4AD, United Kingdom.}

\author{Gustavo P\'{e}rez-Segurana}%
\author{Ian R. Bailey}
\affiliation{%
 Department of Physics, University of Lancaster \& Cockcroft Institute, Bailrigg, Lancaster, LA1 4YB, United Kingdom 
}%

\author{Sara Thorin}
\author{Bill Kyle}
\affiliation{
 MAX IV Laboratory, Lund University, Fotongatan 2, 224 84 Lund, Sweden.
}
\author{Jonas Bj\"orklund Svensson}
\affiliation{
 Department of Physics, Lund University, P.O. Box 118, SE-22100 Lund, Sweden.
}%

\date{\today}

\begin{abstract}
Electron bunch compressors formed of achromat arcs have a natural advantage over the more commonly used chicane compressors in that linearisation of the longitudinal phase space is of the correct sign to compensate for the curvature imprinted by rf acceleration. Here we extend the utility of arc compressors to enable variation of the longitudinal compaction within a fixed footprint. We also show that this variability can be achieved independently order-by-order in momentum deviation. The technique we employ consists of additional dipoles, leading to the advantageous property that variability can be achieved without incurring significant penalty in terms of chromatic degradation. We show this by comparison to an alternative system where additional quadrupoles are utilised to enable variation of momentum compaction. Each of these alternative approaches are being considered in the context of an upgrade of the MAX IV linac, Sweden, to enable a soft X-ray free-electron laser (FEL) in addition to its existing functions.
\end{abstract}

\maketitle

\section{Introduction}
Magnetic bunch compressors are an increasingly common subsystem within linear electron accelerators for high brightness applications. They are used to manipulate the longitudinal phase space of a bunch for which there is a correlation between momentum deviation and longitudinal position caused by a preceding off-crest radio-frequency (rf) acceleration. For example, they are used to reduce the bunch length, and hence increase the peak current, in electron bunches intended to drive a free-electron laser (FEL). Bunch compressors are necessarily momentum dispersive, as the trajectory deviation with momentum is the means by which longitudinal position changes within the bunch are made. Because of this need to transport and control bunches with large momentum spreads, the chromatic properties of bunch compressors are of paramount importance.\par
Bunch compressors are classified by the sign of the momentum compaction they naturally produce at first order. If compression is achieved when the head of the bunch enters the system at a lower energy than the centroid, and the tail of the bunch has higher energy, we label it ``chicane-like'' and combine with pre-acceleration of the bunch on the rising side of the rf waveform. If the opposite is true we label the system ``arc-like'' and combine with pre-acceleration of the bunch on the falling side of the rf waveform.\par
\begin{figure}[b]
    \centering
    \includegraphics[width=0.50\textwidth]{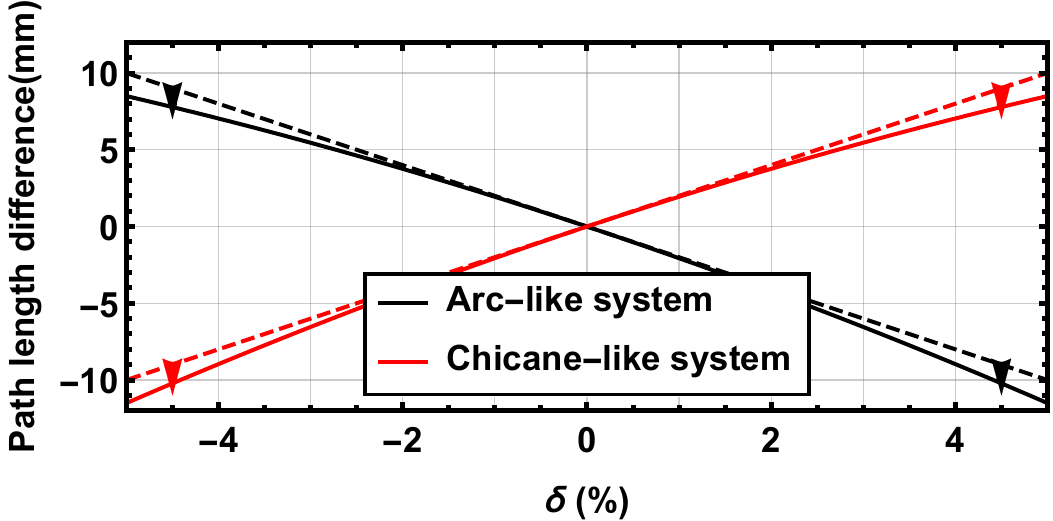}
    \caption{Path length difference with respect to the design trajectory as a function of momentum deviation for uncorrected arc-like (black) and chicane-like (red) compressors showing their respective deviations from linearity (dashed). $R_{56}=-0.2\,$m $T_{566}=-0.6\,$m in the arc-like system and $R_{56}=+0.2\,$m $T_{566}=-0.6\,$m in the chicane-like system.}
    \label{fig:PathLengthChicaneArc}
\end{figure}
\begin{figure}[b]
    \centering
    \includegraphics[width=0.50\textwidth]{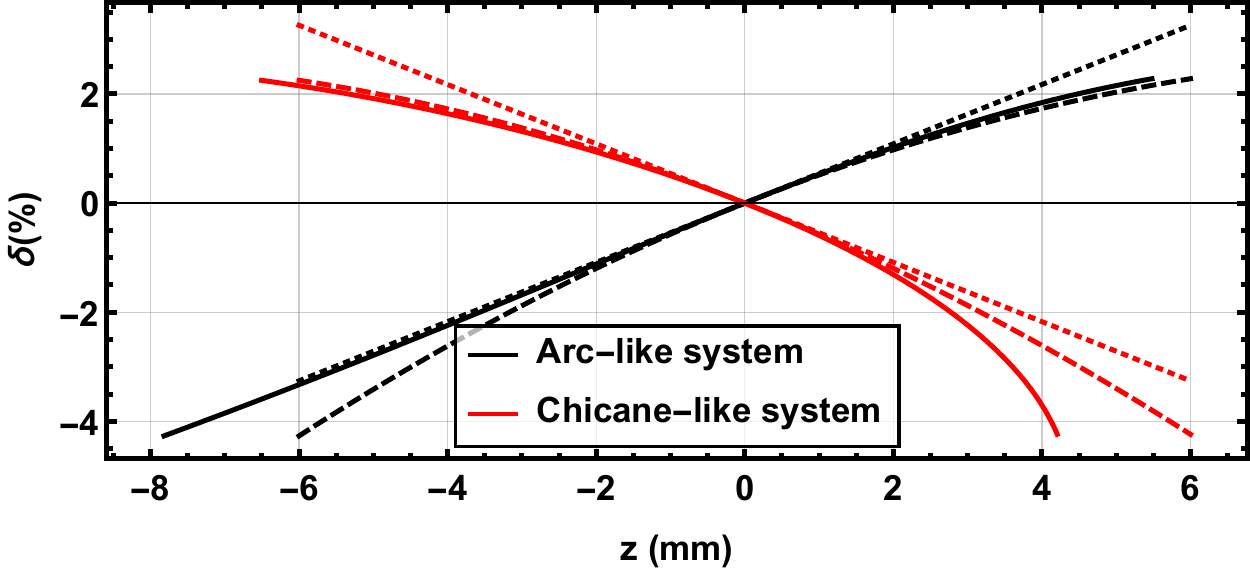}
    \caption{Longitudinal phase space of chirped bunches for compression in an arc-like system (black) and chicane-like (red). Ideal ``linear'' bunch (dotted),  with curvature from the fundamental rf (dashed) and effect of negative $T_{566}$ (solid) shown. We see that the curvature is exacerbated in the chicane-like system, and alleviated in the arc-like system.}
    \label{fig:LPSChicaneArc}
\end{figure}

Using the standard matrix notation for beam transfer~\cite{Brown1982ASpectrometers}, the 6D coordinates $X=(x,x',y,y',z,\delta)$ of a particle in the bunch are transformed with the map
\begin{equation}
\begin{split}
    X_{i,f}=&
    \sum_{j} R_{ij} X_{j}
    + \sum_{jk} T_{ijk} X_{j} X_{k}\\& 
    + \sum_{jkl} U_{ijkl} X_{j} X_{k} X_{l}+ \dots
\end{split}
\end{equation}
where $i$ is the index of the vector $X$ and $\delta=\Delta p/p$ is the momentum deviation from the reference particle. $R_{ij}$, $T_{ijk}$ and $U_{ijkl}$ are the first, second and third order transfer matrices respectively.
The longitudinal position at the end of the bunch compressor as a function of momentum deviation can therefore be expressed as 
\begin{equation}
    z_{f}=z_{i}
    +R_{56}\delta
    +T_{566}\delta^{2}
    +U_{5666}\delta^{3}
    +\mathcal{O}(\delta^{4})
\end{equation}
where $z_{f}$ and $z_{i}$ are the final and initial longitudinal positions of the particle in the bunch, and $R_{56}$, $T_{566}$ and $U_{5666}$ are also known as the first, second and third order longitudinal dispersions and particles at the head of the bunch have $z>0$. In this description, a chicane-like system has a negative $R_{56}$ and an arc-like system positive $R_{56}$ (for codes that utilise $t$ instead of $z$ as the longitudinal coordinate the opposite is true, hence our preference for the ``like'' descriptions as they are unambiguous). A consideration of the second order longitudinal dispersion immediately reveals a disadvantage of the more common chicane-like implementations~\cite{Thorin2010BunchInjector,Owen2012ALinacs}; both arc- and chicane-like systems have naturally negative $T_{566}$. To see this one may compare the path length as a function of energy deviation for both systems, shown in Fig.~\ref{fig:PathLengthChicaneArc}. At nominal energy one sees that the linear slopes are of opposite sign, but the curvature is of the same sign. In an arc-like system this property is welcome as the natural rf curvature (which places both head and tail at more positive $z$ than a linear, falling rf chirp), is partially cancelled by the effect of $T_{566}$. Whereas in a chicane-like system, the $T_{566}$ exacerbates the curvature of the rising rf chirp, necessitating deliberate cancellation by an additional external system. Typically this takes the form of a higher-harmonic rf linearising cavity, which requires an expensive, additional rf system as well as facility space\footnote{Linearisation can also be achieved with strong linac wakefields, as in the FERMI facility~\cite{DiMitri2014ElectronFree-electron-lasers} where the final linac sections are backward travelling wave structures. This lessens the correction required from a higher harmonic system, but does not eliminate it as tuning of the linearisation is required. The arc-like systems presented here include such tune-ability.}. We illustrate the contrasting situations in Fig.~\ref{fig:LPSChicaneArc}.\par
\begin{figure*}
    \centering
    \includegraphics[width=0.95\textwidth]{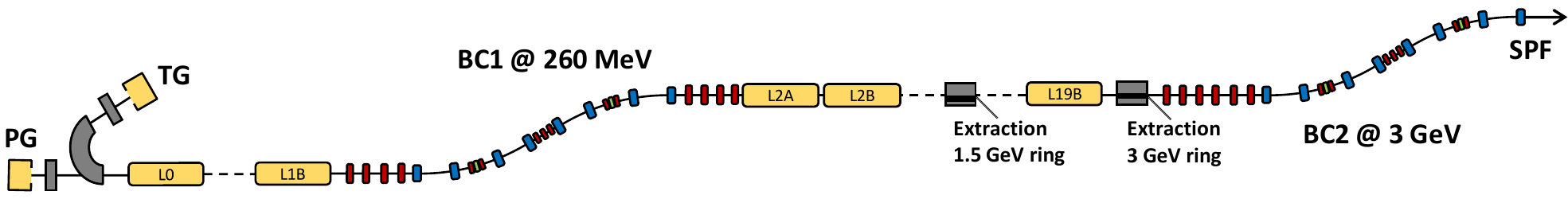}
    \caption{Layout of the MAX IV linac. Bunch compressor 1 (BC1) is located at $260$\,MeV. Extractions to the $1.5$ and $3.0$\,GeV rings are at the exit of acceleration sections $2$ \& $3$ respectively. BC2 is located at $3.0$\,GeV and feeds a short-pulse facility (SPF) and future soft X-ray FEL.}
    \label{fig:max_iv_layout}
\end{figure*}

This context is particularly relevant for the MAX IV facility~\cite{Martensson2018TheSource}. MAX IV is primarily a near diffraction-limited (or fourth generation) storage ring facility utilising full energy injection from an S-band linac~\cite{Thorin2014TheLinac} into two rings, one at $1.5\,$GeV and a second at $3\,$GeV. It was foreseen at the design stage that this linac could also drive spontaneous undulator radiation in a Short Pulse Facility (SPF), and be adapted in the future to drive a soft X-ray FEL. In order to generate high peak current, longitudinal phase space linearisation is necessary for both applications. The implementation of arc-like compressors achieves this without the need for expensive harmonic rf, the disadvantages being the transverse displacement of the beamline sections on either side of each compressor and the fixed $R_{56}$. The location of the compressors within the linac are shown in Fig.~\ref{fig:max_iv_layout}.\par
In addition to the longitudinal properties, when transporting and manipulating large energy spread bunches one must be mindful of the transverse chromatic properties. In order to analyse this in the various systems we study the {\it chromatic amplitude} function~\cite{Montague1979LinearCorrection,Lindstrm2016DesignLines}
\begin{equation}
    W\equiv\sqrt{\left(
    \frac{\partial\alpha}{\partial\delta}-\frac{\alpha}{\beta}\frac{\partial\beta}{\partial\delta}
    \right)^{2}
    +\left(
    \frac{1}{\beta}\frac{\partial\beta}{\partial\delta}
    \right)^{2}}\,,
\end{equation}
which quantifies the linear chromatic error of our beam as a distribution and thus the energy dependence of the focusing. Lack of control over the chromatic derivatives of the Twiss parameters $\partial\alpha/\partial\delta$ and $\partial\beta/\partial\delta$ returns an increased projected and sliced emittance as a consequence of the disparities in dispersion and focusing strengths for different beam slices. The property $W=0$ is denominated apochromaticity. It is worth noting that we do not use the single-particle variable {\it chromaticity}, this is more useful in storage rings where fractional energy deviations of order $10^{-5}$ are typical. In linacs operated off-crest, we expect energy spreads at the percent level in the low energy sections of the machine, therefore a measure based on energy variation within the particle bunch is far preferable. \par
\begin{figure}[t]
    \centering
    \includegraphics[width=0.50\textwidth]{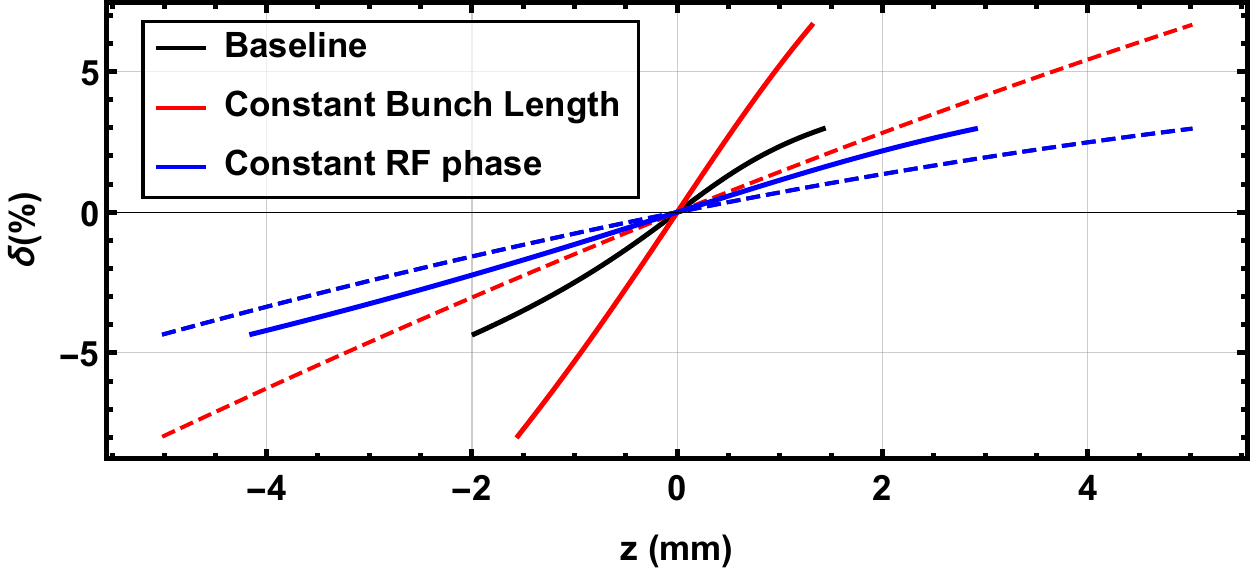}
    \caption{Longitudinal phase space of chirped bunches before (dashed) and after (solid) compression. The baseline bunch (black) is accelerated 20$^{\circ}$ on the falling side of the crest (overlapped by dashed blue line) is compressed with $R_{56}=-0.1\,$m and linearised with $T_{566}=-0.70\,$m. After reducing $R_{56}$ to $-0.05\,$m, a bunch of equal initial length (red) requires acceleration $\sim 40^{\circ}$ off-crest with $T_{566}=-0.08\,$m. If the rf phase is kept constant (blue), the final bunch length increases and the $T_{566}$ required for linearisation remains unchanged.}
    \label{fig:LPSVariableR56Cases}
\end{figure}

The MAX IV storage rings and SPF were completed in 2016 and approval has been given for a detailed design of a soft X-ray FEL~\cite{Andersen2017TheIV}. Experience at FEL facilities in the interim period has highlighted the importance of variability in the $R_{56}$ of the compressors. The additional degree of freedom in the longitudinal phase space allows either manipulation of the bunch length at a fixed rf phase, or manipulation of the rf phase (and therefore energy spread) at constant bunch length. These two cases and their implications on linearisation are shown in Fig.~\ref{fig:LPSVariableR56Cases}. At a constant rf phase, the $T_{566}$ required for linearisation remains the same. If we instead alter the rf phase keeping a constant bunch length the linearising $T_{566}$ varies as~\cite{Floettmann2001GenerationSpace}
\begin{equation}
    T_{566}=\frac{E_{0}^{2}}{2 k V_{RF}^{2}\textrm{tan}^{3}\phi}\, .
\end{equation}
\par
Motivated by this, we set out in this work two options to ``retrofit" variability of $R_{56}$ into the existing layout of MAX IV. We term these the {\it additional quadrupole} and {\it additional dipole} solutions, respectively. In doing so we establish the additional dipole solution as having the advantages of apochromaticity and order-by-order control of the momentum compaction, the disadvantage being that part of the trajectory must have a variable horizontal position.\par
Although the motivation is to introduce flexibility into the MAX IV facility, our observations on these arc-like compressors are generic and have wide potential future applicability in situations where large energy spread beams must be longitudinally manipulated. In particular, any laser (LWFA) or plasma (PWFA) wakefield generated beam would benefit from the deployment of the following proposed schemes for transport and conditioning.

\section{Existing Fixed $R_{56}$ Arc-Like Compressors}
\begin{figure}[ht]
    \centering
    \includegraphics[width=0.50\textwidth]{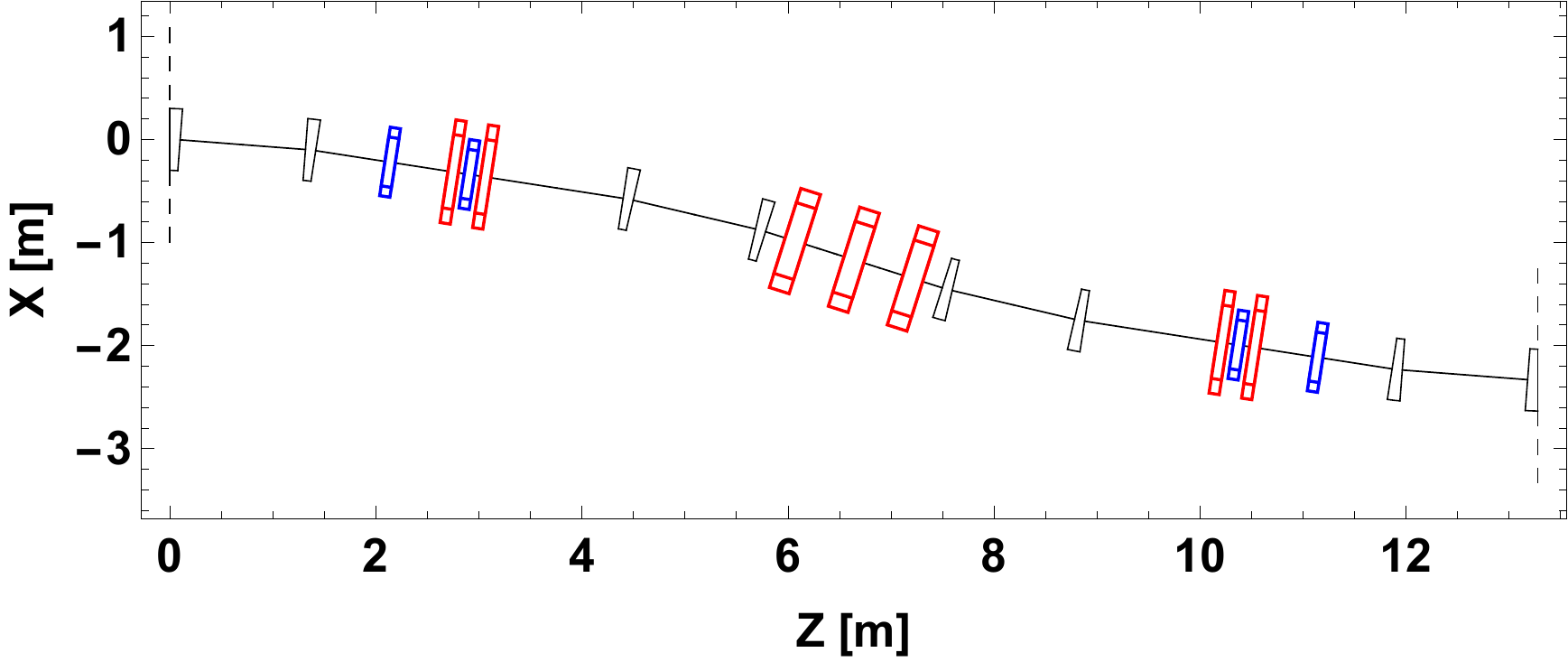}
    \caption{Existing MAX IV Bunch Compressor 1. Dipoles in black, quadrupoles in red and sextupoles in blue.}
    \label{fig:OriginalLayout}
\end{figure}

\begin{figure}[ht]
     \centering
     \begin{subfigure}[h]{0.50\textwidth}
         \centering
         \includegraphics[width=\textwidth]{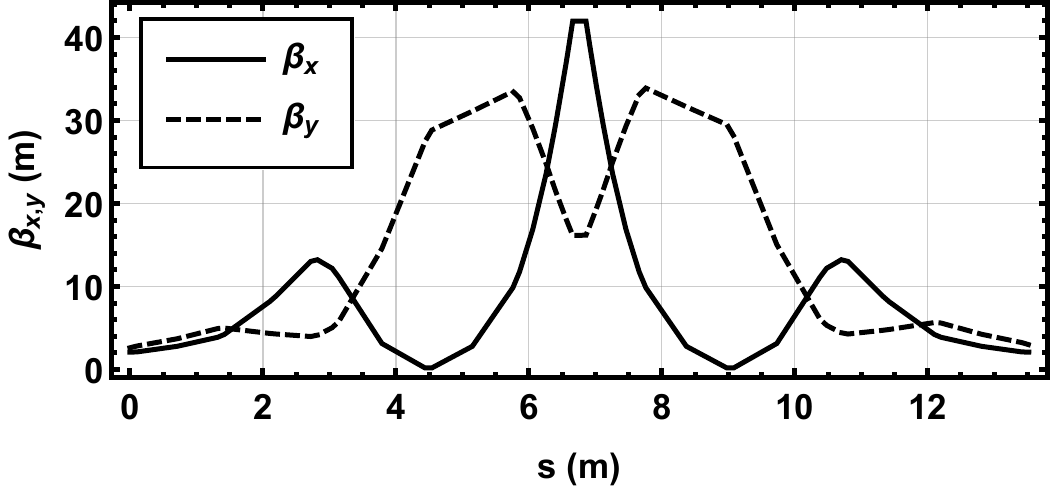}
     \end{subfigure}
     \\
     \begin{subfigure}[h]{0.50\textwidth}
         \centering
         \includegraphics[width=\textwidth]{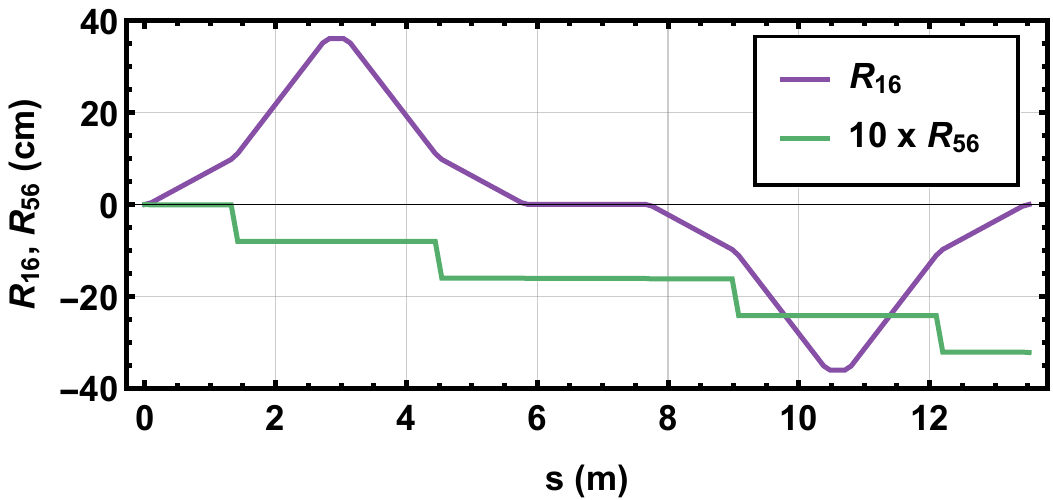}
     \end{subfigure}
     \\
     \begin{subfigure}[h]{0.50\textwidth}
         \centering
         \includegraphics[width=\textwidth]{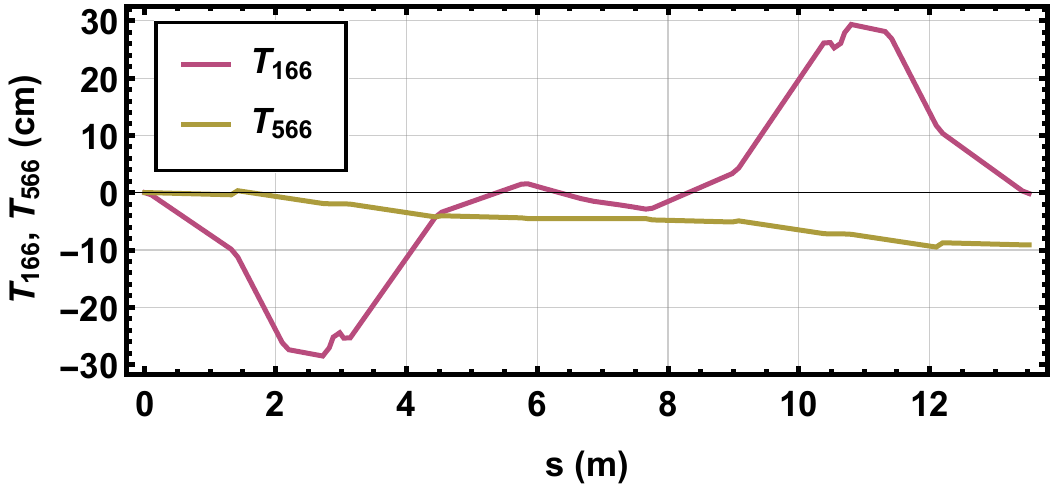}
     \end{subfigure}
     \\
     \begin{subfigure}[h]{0.50\textwidth}
         \centering
         \includegraphics[width=\textwidth]{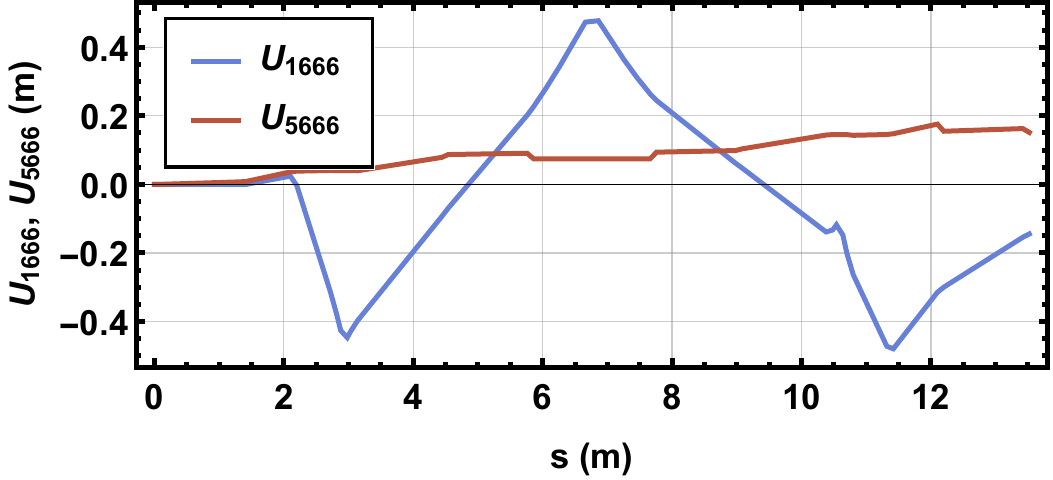}
     \end{subfigure}     
        \caption{Optics and first, second and third order horizontal and longitudinal dispersion functions of existing MAX IV BC1.}
        \label{fig:JonasOptics}
\end{figure}

The two bunch compressors at MAX IV each consist of two back-to-back double-bend achromats with three phase advance matching quadrupoles separating them. The centre of each arc hosts a pair of quadrupoles to enforce the first-order achromatic condition and one sextupole for varying the second-order longitudinal momentum compaction away from the naturally over-linearising value that is apparent in Fig.~\ref{fig:LPSChicaneArc}. The first compressor (BC1) is shown in Fig.~\ref{fig:OriginalLayout}. It is located $\sim 18\,$m from the cathode at beam energy $\sim 275\,$MeV. The second compressor (BC2) is of optically similar design and is situated $\sim 298\,$m from the cathode at beam energy $\sim 3000\,$MeV. In both systems, the beamline central trajectory is offset transversely by $\sim 2\,$m. For the rest of this paper we work with the first bunch compressor. The design optics and dispersion function for BC1 are shown in Fig.~\ref{fig:JonasOptics}.

In the initial design shown, the sextupoles are set such that the $T_{566}$ is reduced from nominal. The resulting $R_{56}$ and $T_{566}$ of this lattice are $-32\,$mm and $-91\,$mm respectively. A recent paper described a method of inserting sextupoles to ensure achromaticity of this design to third order~\cite{BjorklundSvensson2019Third-orderBeams}. We will comment on the ease of replicating this feature in the variable $R_{56}$ proposals that follow.

\section{Additional Quadrupole Variable $R_{56}$ Compressor}
\begin{figure}[h]
    \centering
    \includegraphics[width=0.50\textwidth]{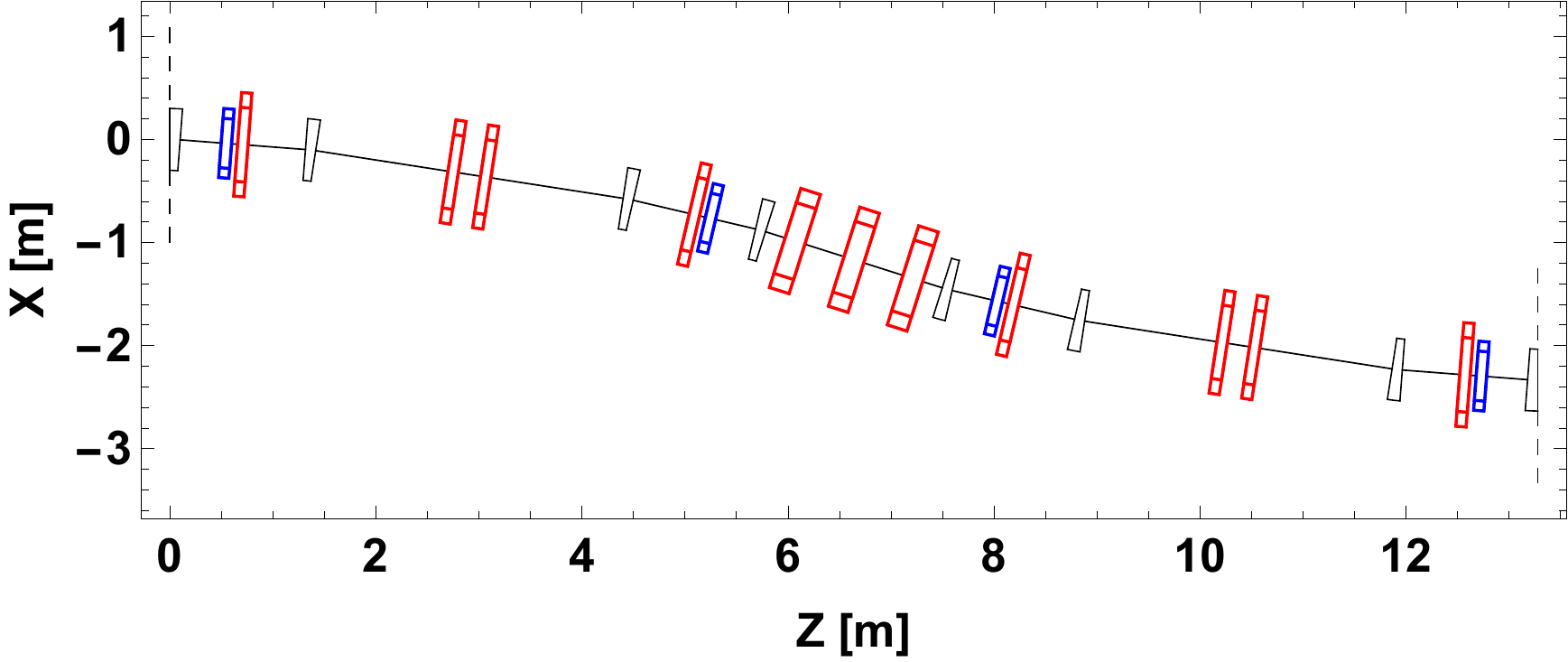}
    \caption{Additional quadrupoles layout. Dipoles in black, quadrupoles in red and sextupoles in blue.}
    \label{fig:ExtraQuads_Layout}
\end{figure}

\begin{figure}[ht]
     \centering
     \begin{subfigure}[h]{0.50\textwidth}
         \centering
         \includegraphics[width=\textwidth]{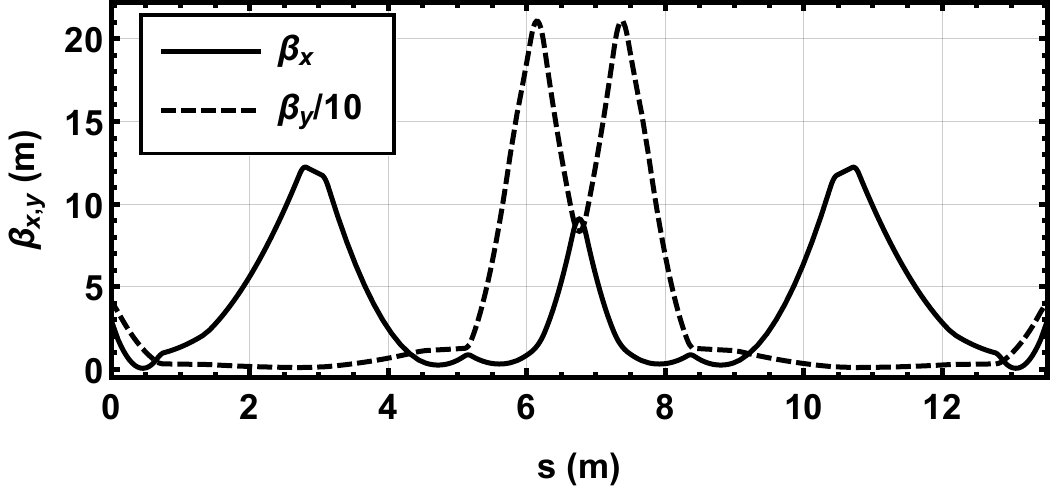}
     \end{subfigure}
     \\
     \begin{subfigure}[h]{0.50\textwidth}
         \centering
         \includegraphics[width=\textwidth]{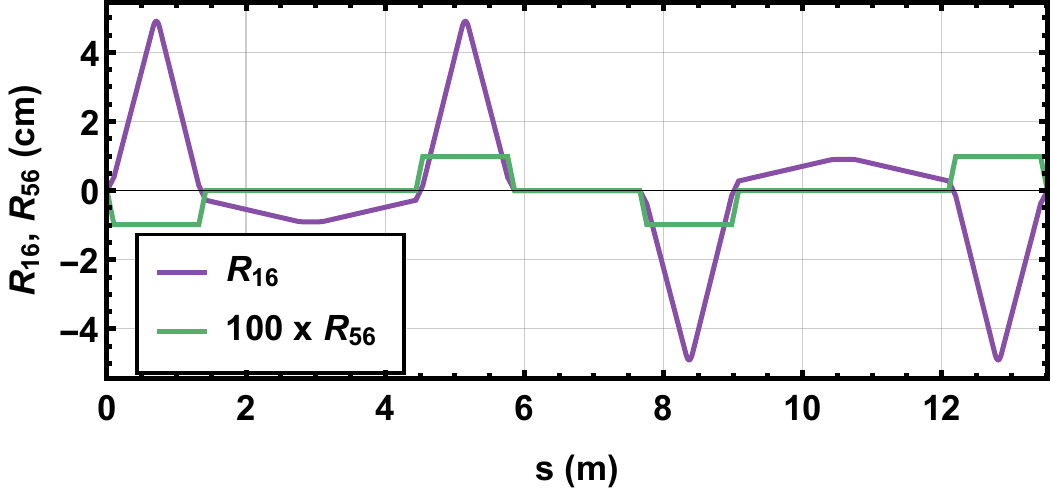}
     \end{subfigure}
     \\
     \begin{subfigure}[h]{0.50\textwidth}
         \centering
         \includegraphics[width=\textwidth]{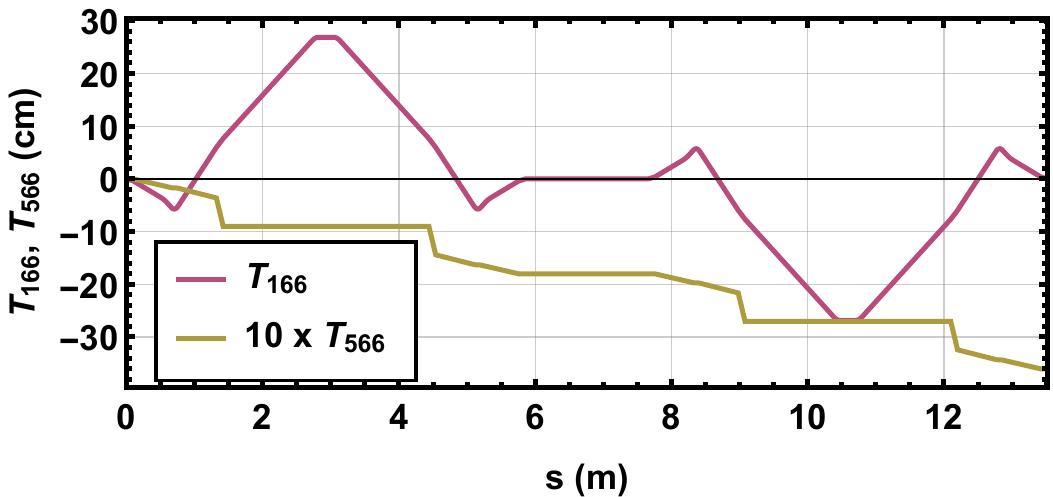}
     \end{subfigure}
     \\
     \begin{subfigure}[h]{0.50\textwidth}
         \centering
         \includegraphics[width=\textwidth]{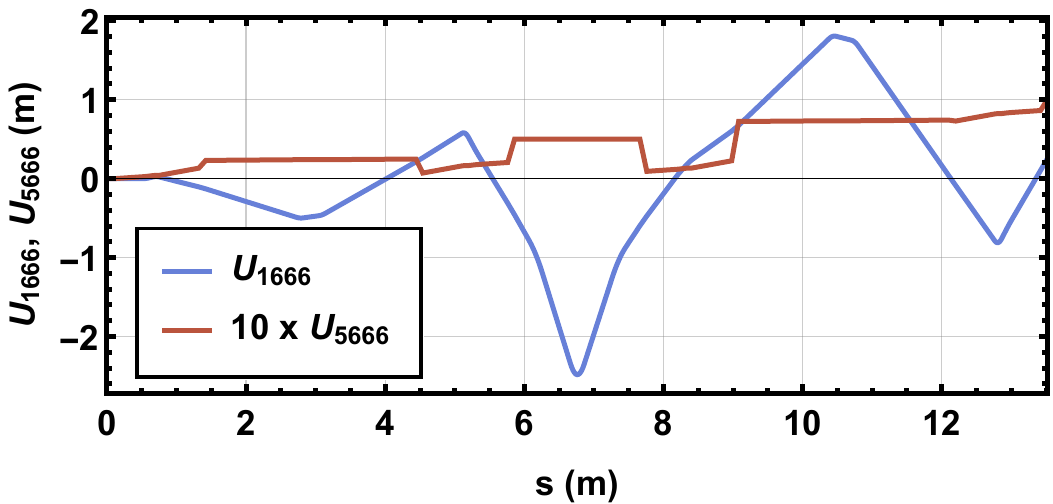}
     \end{subfigure}     
        \caption{Optics, first, second and third order longitudinal dispersion functions of additional quadrupole solution tuned to isochronous condition.}
        \label{fig:ExtraQuadOptics}
\end{figure}

The most conceptually simple alteration one can make to the original compressors that enables manipulation of the $R_{56}$ is the insertion of additional quadrupoles between each pair of dipoles as shown in Fig.~\ref{fig:ExtraQuads_Layout}.
Varying the strength of these quadrupoles changes the value of the dispersion as it reaches the following dipole, thereby changing the $R_{56}$ of each achromat generated via the well-known expression,
\begin{equation}
    R_{56}=-\int\frac{R_{16}}{\rho} \,ds \,,
\end{equation}
where $\rho$ is the bending radius of the design trajectory and $R_{16}$ is equivalent to the linear horizontal dispersion. The quadrupoles at the centre of each achromat are tuned to symmetrize the dispersion at each side. Following this scheme we can reduce the final $R_{56}$ magnitude continuously and even drive it to positive values. An example tuning is shown in Fig.~\ref{fig:ExtraQuadOptics} where we choose the isochronous condition. It should be noted that the quadrupole $k$-value for this tuning is rather large, at $\sim 33\,$m$^{-2}$.
\par
\begin{figure}
    \centering
    \includegraphics[width=0.50\textwidth]{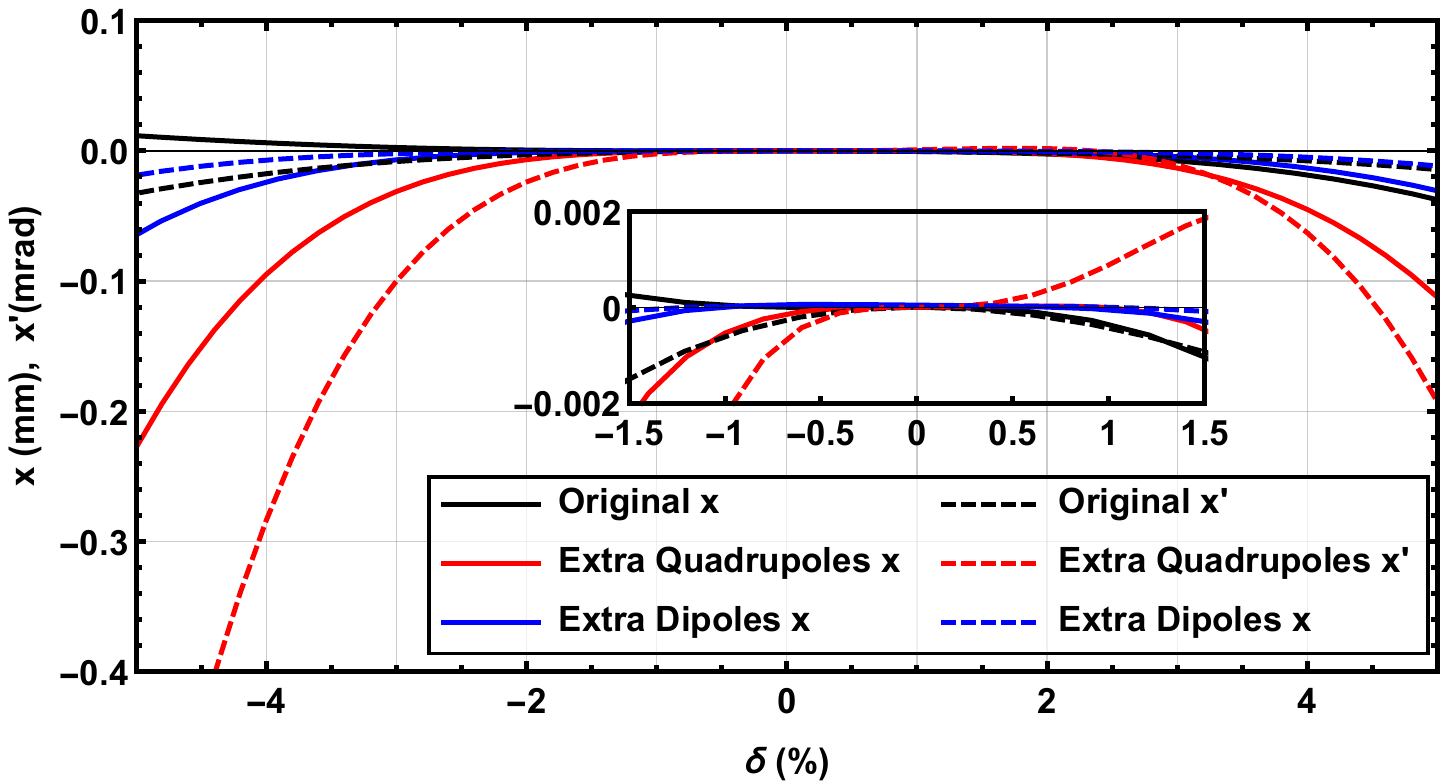}
    \caption{Horizontal momentum acceptance of the original, fixed design with chromatic corrections (black), variable with extra quadrupoles (red) and variable with extra dipoles (blue) solutions.}
    \label{fig:ExtraQuadsMomAcceptance}
\end{figure}
\begin{figure}
    \centering
    \includegraphics[width=0.50\textwidth]{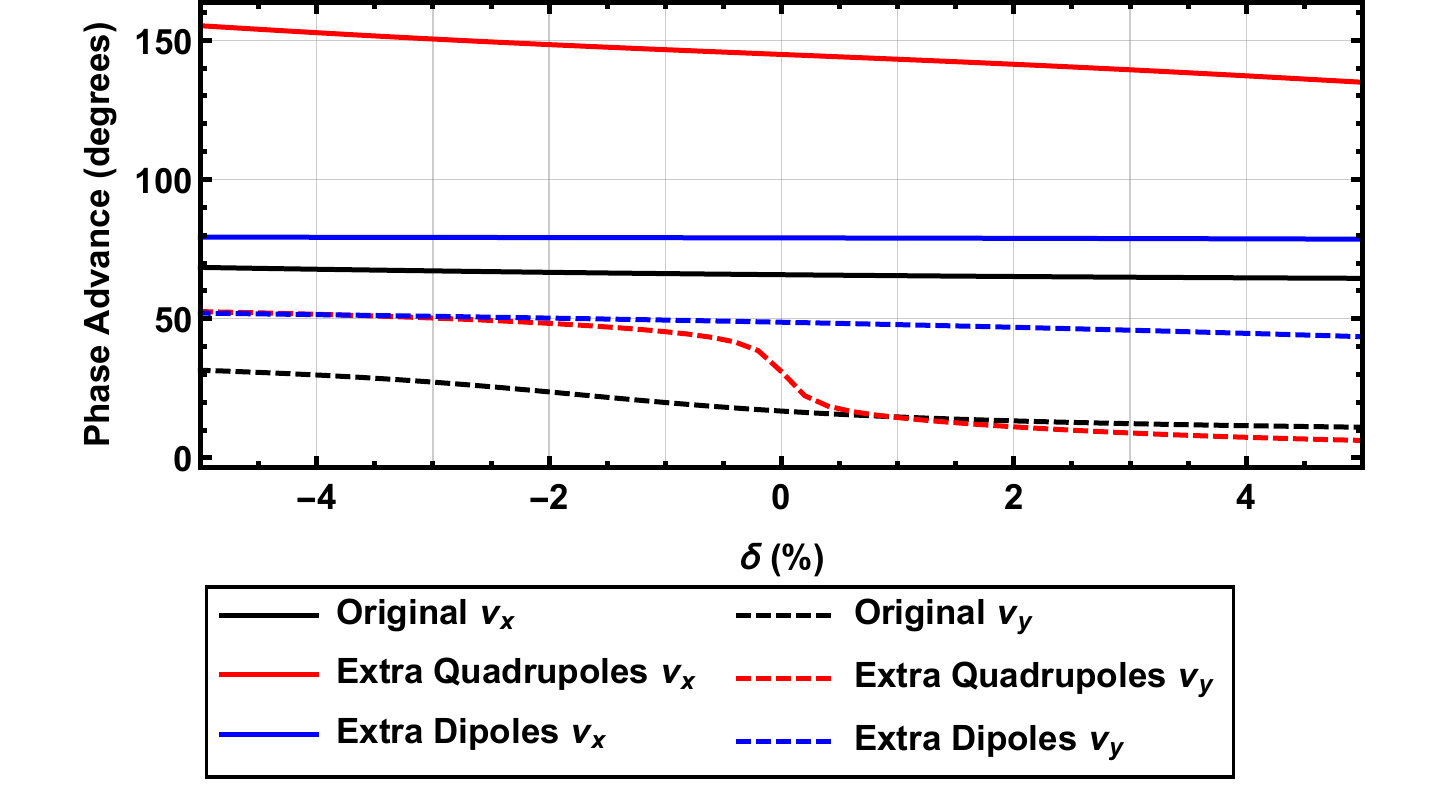}
    \caption{Phase advance dependence on $\delta$ at the end of BC1 for the Original lattice (black), extra quadrupoles (red) and extra dipoles solutions (blue)}
    \label{fig:Chromaticity}
\end{figure}
\begin{figure}
    \centering
    \includegraphics[width=0.50\textwidth]{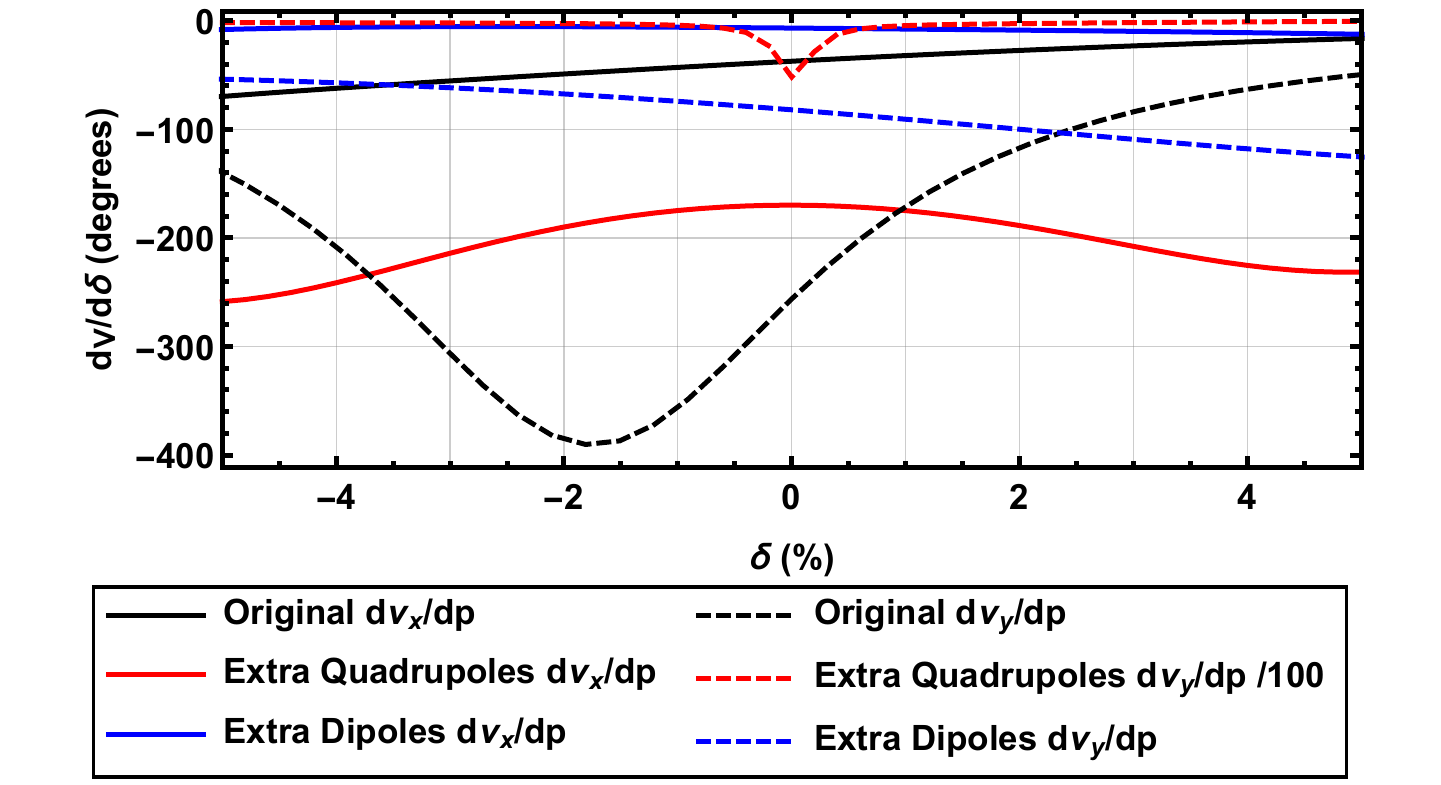}
    \caption{Derivative of the phase advance with respect to $\delta$ at the end of BC1 for the original lattice (black), extra quadrupoles (red) and extra dipoles solution (blue).}
    \label{fig:Chromaticity2}
\end{figure}

In order to minimise second order horizontal dispersion, $T_{166}$, we also insert additional sextupoles at the location where dispersion is highest for all the range of $R_{56}$ configurations, following~\cite{BjorklundSvensson2019Third-orderBeams}. This is naturally reduced in comparison to the initial design as the second order longitudinal dispersion across the whole bunch compressor is related to the transverse dispersion via~\cite{Robin1993Quasi-isochronousRings}\footnote{Care should be taken when deriving this expression. Premature truncation of the power series expansions of the coordinates results in an erroneous additional $R_{16}$ term. This cancels exactly on retention of the next highest order terms.}
\begin{equation}
    T_{566}=-\int
    \left[
    \frac{R_{26}^{2}}{2} +
    \frac{T_{166}}{\rho}
    \right] \, ds.
\end{equation}
Reasonable momentum acceptance for $\delta\leq 3\%$ is maintained, as shown in Fig.~\ref{fig:ExtraQuadsMomAcceptance}. However since all quadrupoles used for the dispersion manipulation are focusing, the chromatic performance is harmed. The chromatic behaviour in the vertical axis is also spoiled by the strong focusing as shown by the phase advance response to deviations in momentum (Figs.~\ref{fig:Chromaticity} and \ref{fig:Chromaticity2}).


\section{Additional Dipole Variable $R_{56}$ Compressor}
A standard solution adopted by many electron accelerators driving FELs is the {\it variable} four-dipole chicane bunch compressor~\cite{Borland2000AFEL}. The technique adopted is a mechanical change in the centroid trajectory, achieved through moving dipoles and beam tubes mounted on an actuated table. Here we adapt that approach in the context of arc-like compressors, analogously relaxing the concept of a fixed trajectory in order to gain advantage in terms of chromatic behaviour over the strongly-focusing additional quadrupole solution.\par
Our technique draws inspiration from the recent deployment of ``anti-bend" cells in low emittance storage ring lattice designs~\cite{Streun2014TheLattices,Riemann2019LowBends}, where the purpose is also the control of dispersion.\par
\begin{figure}[h]
    \centering
    \includegraphics[width=0.50\textwidth]{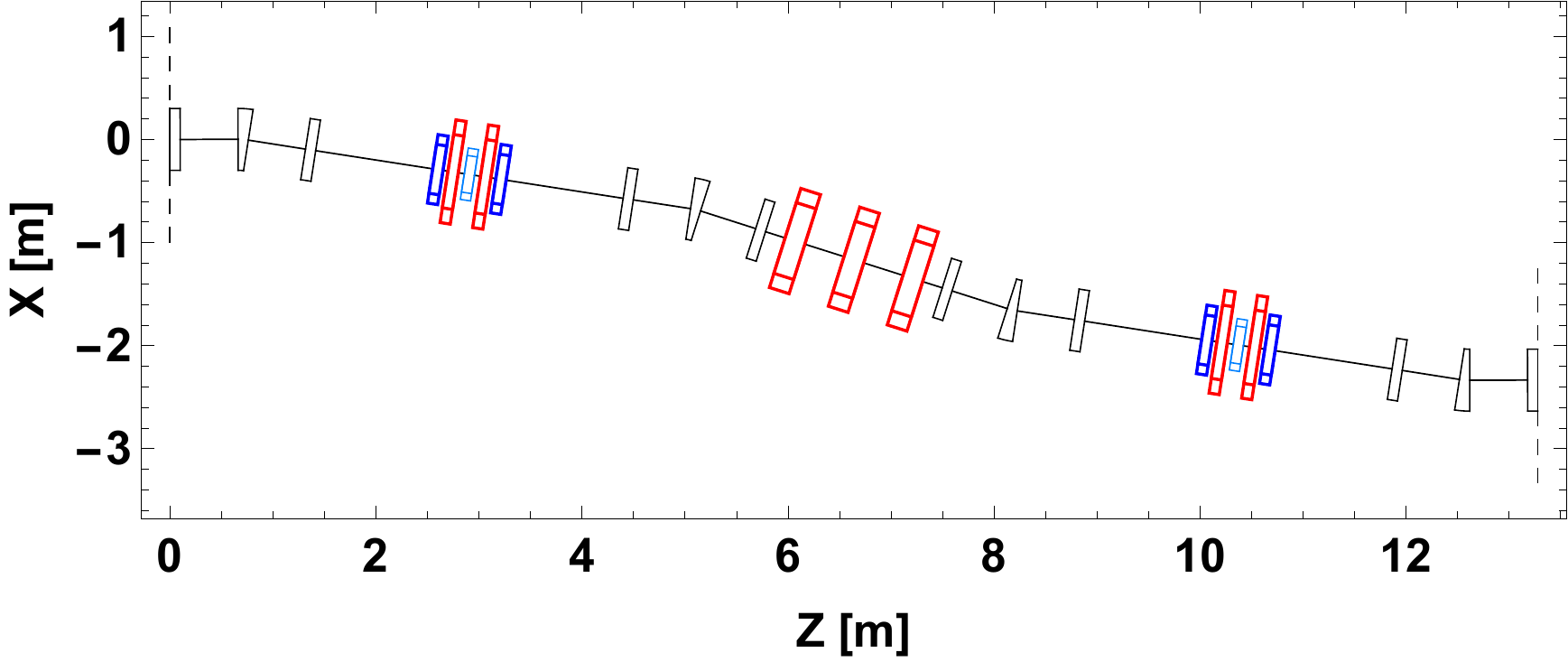}
    \caption{Additional Bends layout. Dipoles in black, quadrupoles in red, sextupoles in blue and octupoles in light blue.}
    \label{fig:ExtraBendLayout}
\end{figure}
\begin{figure}[h]
    \centering
    \includegraphics[width=0.50\textwidth]{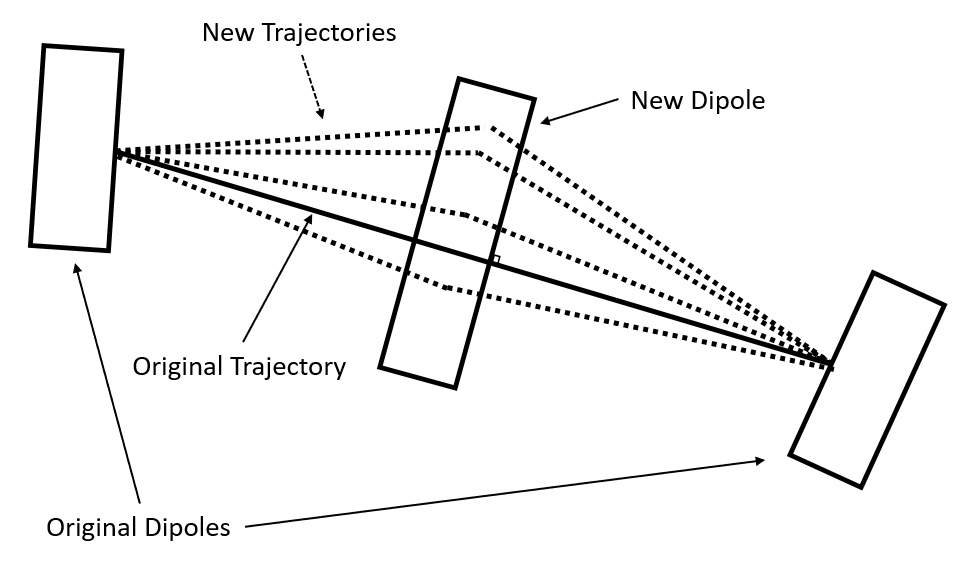}
    \caption{Schematic of new dipole with the original trajectory (solid) and different new trajectories (dashed) for different $R_{56}$ values. For very low values of $R_{56}$ the original dipoles become reverse bends.}
    \label{fig:ExtraDipoleSketch}
\end{figure}

We insert an additional dipole in between each pair of existing ones. The resulting layout for MAX IV BC1 is shown in Fig.~\ref{fig:ExtraBendLayout}. As we are working within a fixed footprint, the total bending angle of each three-dipole subsystem must remain constant. Thus the bend angle of our additional dipole is set by alterations of the strength of the existing two dipoles as shown in Fig.~\ref{fig:ExtraDipoleSketch}.\par
Unlike a four-dipole compressor, the excursion of the trajectory as the $R_{56}$ is varied is small enough that no actuation is needed, merely an expanded beam tube. However, as the trajectory explores a large range of offsets within the new dipole, care must be given in considering the properties of the good field region. The original dipoles are of combined function in order to provide additional vertical focusing, ensuring the two planes are not very different in terms of phase advance over the system. They share $\theta:k_{1}:k_{2}$ ratios with the initial parameters $\pm76\,\textrm{mrad}:-3.0\,\textrm{m}^{-2}:3.1\,\textrm{m}^{-3}$. Since different $R_{56}$ configurations have different trajectories across these new dipoles, designing a magnet with specific quadrupole and sextupole components for each trajectory is challenging. Therefore it was decided to use flat dipoles (no quadrupole or sextupole moments), leaving open the option to implement them with the correct sign further along the design process if it became necessary to relax the optics or sextupole strengths.
Additionally, we considered the magnet edge angles, as these have focusing effects in both vertical and horizontal planes. The existing dipoles are rectangular magnets, such that $e_{1} + e_{2} =\theta$ where $e_{1}$ and $e_{2}$ are the entry and exit edge angles and $\theta$ is the total bending angle. Here we consider 3 different configurations.
Firstly, rotate the dipoles for each desired $R_{56}$ configuration, such that entry and exit angles are always equal. Secondly, a simpler option in terms of implementation is to set $e_{1}=0$ and $e_{2} = \theta$. Finally, we consider a new design of a magnet with curved edges such that $e_{1} = e_{2} = 0$ for all $\theta$. We want to minimise horizontal focusing as it would counteract our dispersion generation. Both stationary and rotating dipoles configurations give a net zero horizontal focusing, while the sector magnet provides a positive horizontal focus. Therefore we select the stationary dipole configuration as the difference in vertical focus is minimal ($\propto \theta^{3}$), and it is mechanically simpler to implement since maintaining the balance of the edge angles would require rotating dipoles. The new magnets are also rectangular and as they are situated in a symmetric point between the original dipoles, their edge angles are $e_{1} = e_{2} = \theta / 2$.

The limits of this strategy lie in the maximum dipole field we can achieve without requiring permanent or superconducting magnets and the design of the new dipoles that must accommodate a wide horizontal acceptance for the different $R_{56}$ configurations. The requirements on the dipole strength can be lessened by using a longer dipole, provided there is space in the beamline. In the original, fixed design, $R_{56} = -32\,\textrm{mm}$.
For $R_{56} = -15\,\textrm{mm}$ we need a horizontal acceptance of $3.1\,$cm, for $R_{56} = 0\,\textrm{mm}$, the acceptance grows to $4.4\,$cm, and finally for $R_{56} = +5\,\textrm{mm}$ the deviation of the centroid trajectory at the entry to the extra dipole with respect to the original design is $4.8\,$cm.

As in the additional quadrupole solution, the quadrupoles in the centre of each achromat are tuned to symmetrize the dispersion at each side.

\begin{figure}[h]
     \centering
     \begin{subfigure}[h]{0.50\textwidth}
         \centering
         \includegraphics[width=\textwidth]{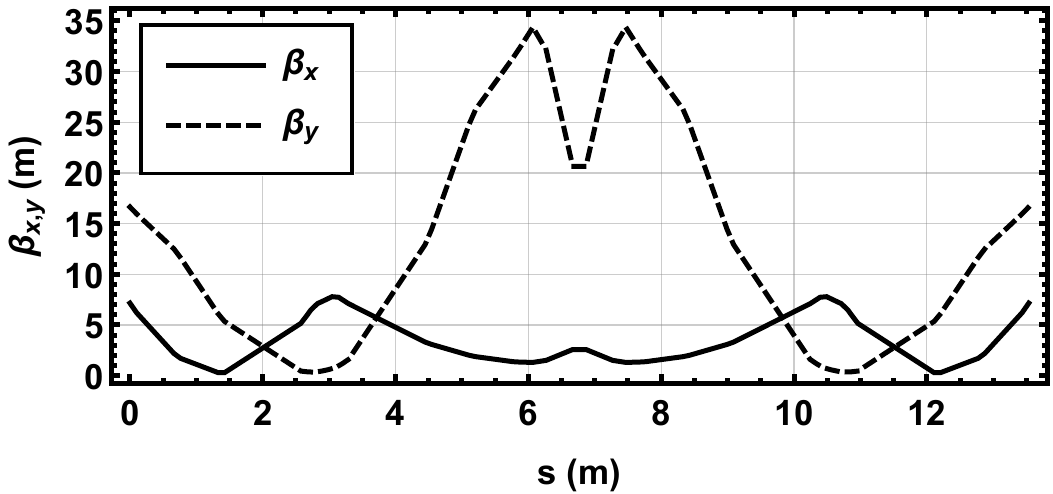}
     \end{subfigure}
     \\
     \begin{subfigure}[h]{0.50\textwidth}
         \centering
         \includegraphics[width=\textwidth]{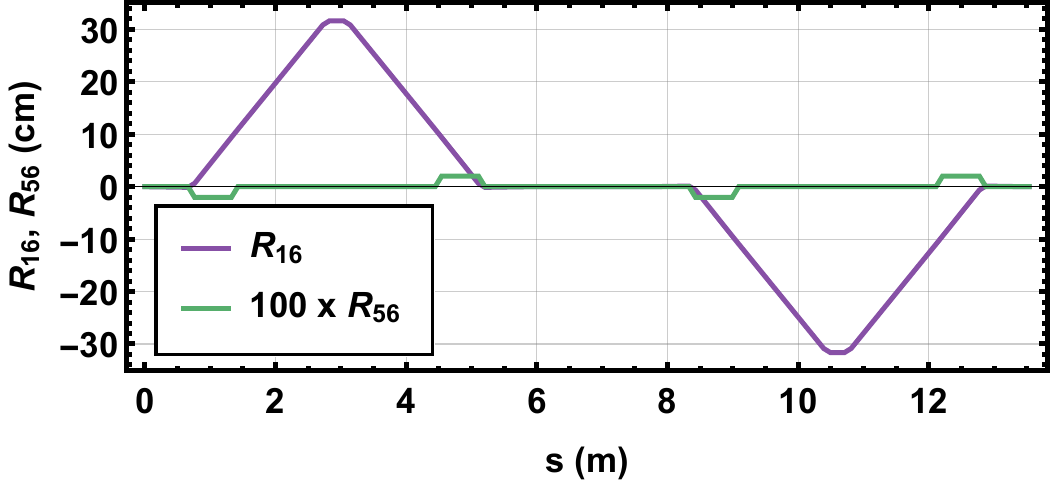}
     \end{subfigure}
     \\
     \begin{subfigure}[h]{0.50\textwidth}
         \centering
         \includegraphics[width=\textwidth]{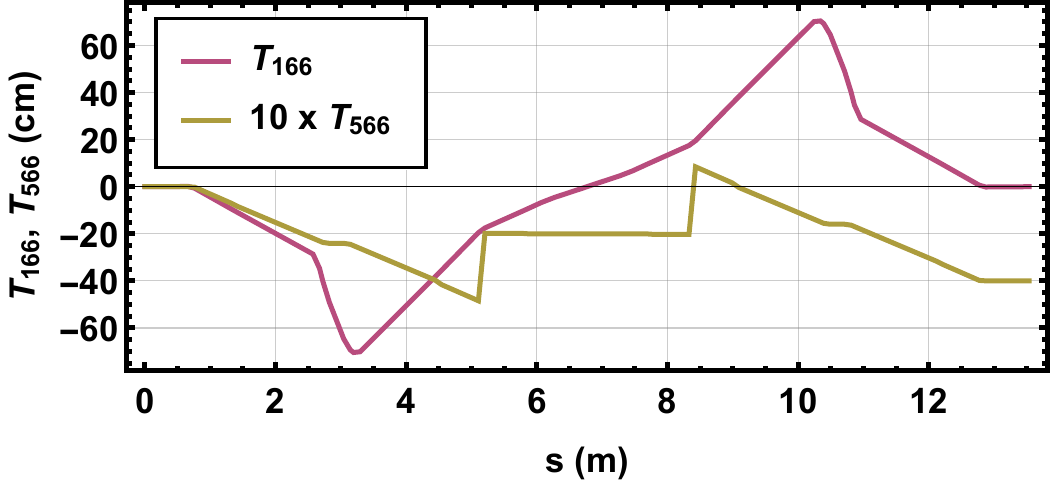}
     \end{subfigure}
     \\
     \begin{subfigure}[h]{0.50\textwidth}
         \centering
         \includegraphics[width=\textwidth]{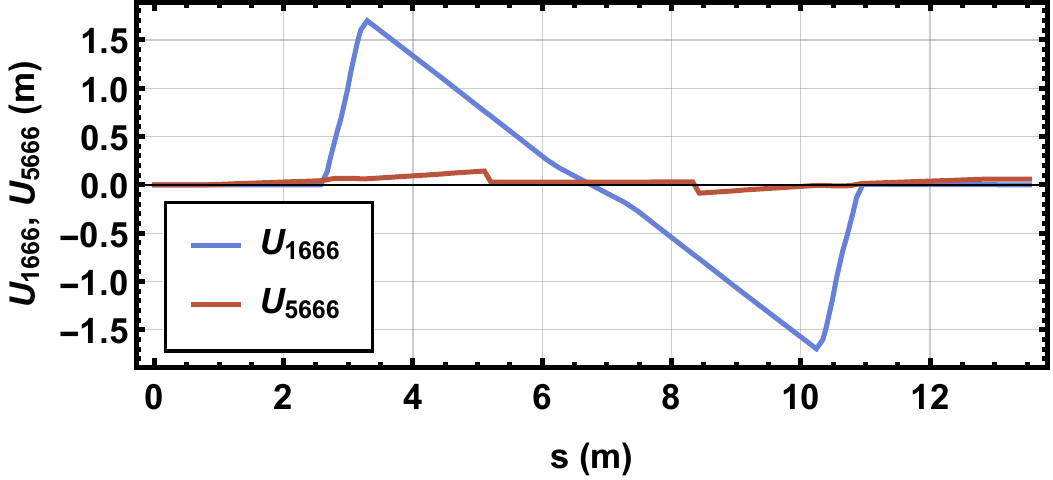}
     \end{subfigure}     
        \caption{Optics and first, second and third order dispersion functions of isochronous MAX IV BC1 with additional dipoles. $R_{56}=0\,$cm $T_{566}=-4.0\,$cm}
        \label{fig:ExtraBendOptics}
\end{figure}

Sextupoles are used to cancel the elements $T_{166}$ (as shown in Fig.~\ref{fig:ExtraBendOptics}) and $T_{266}$ at the end of the bunch compressor, thereby improving its chromatic behaviour. With an additional family of sextupoles we can regain control of the $T_{566}$ term, thus our system exhibits independently tune-able $R_{56}$ and $T_{566}$. Finally, a single pair of octupoles is used to cancel the third order horizontal dispersion, $U_{1666}$, using the anti-symmetry in the bending angles over the whole bunch compressor. A configuration with $R_{56}=0\,$cm and $T_{566}$ shifted from the natural $\sim-9.6\,$cm to $-4.0\,$cm is shown in Fig.~\ref{fig:ExtraBendOptics}. The limits on the $T_{566}$ variability depend on the starting $T_{566}$ for the chosen $R_{56}$, sextupole and octupole strength.

\begin{table}[h]
\caption{Magnet parameters for MAX IV BC1 with additional dipoles. We choose as an example the first-order isochronous, second order linearised solution: $R_{56}=0\,$cm $T_{566}=-4.0\,$cm}
\begin{tabular}{c|ccc}
\hline\hline
                         & $k_{1} (\textrm{m}^{-1})$ & $k_{2} (\textrm{m}^{-2})$ & $k_{3} (\textrm{m}^{-3})$ \\\hline
Dipoles 1, 3, 4 \& 6     & 0.0783                 & -0.08185             & 0                         \\
Dipoles 2 \& 5           & 0                         & 0                         & 0                         \\
Dipoles 7, 9, 10, 12     & 0.0783                 & -0.08185             & 0                         \\
Dipoles 8 \& 11          & 0                         & 0                         & 0                         \\
Quadrupoles 1, 2, 6 \& 7 & 4.88141                   & 0                         & 0                         \\
Quadrupoles 3 \& 5       & -3.1747                   & 0                         & 0                         \\
Quadrupole 4             & 6.30158                   & 0                         & 0                         \\
Sextupole 1              & 0                         & 215.345                   & 0                         \\
Sextupole 2              & 0                         & -116.801                  & 0                         \\
Sextupole 3              & 0                         & 116.801                   & 0                         \\
Sextupole 4              & 0                         & -215.345                  & 0                         \\
Octupoles 1 \& 2         & 0                         & 0                         & -2373.73\\\hline\hline  
\end{tabular}
\end{table}

\begin{figure}[h]
     \centering
     \begin{subfigure}[h]{0.50\textwidth}
         \centering
         \includegraphics[width=\textwidth]{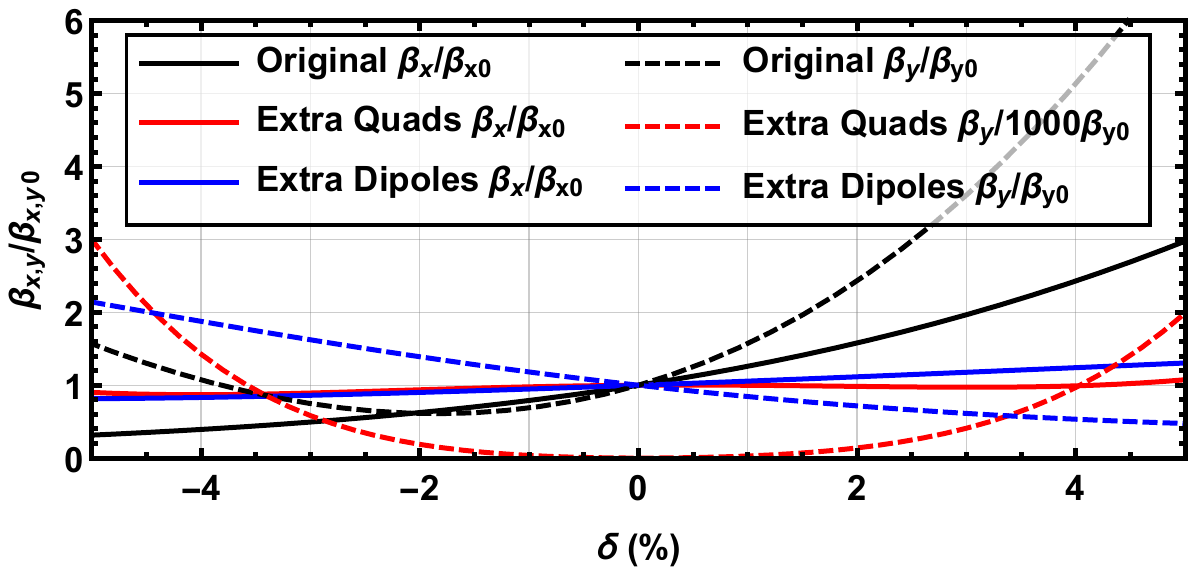}
     \end{subfigure}
     \\
     \begin{subfigure}[h]{0.50\textwidth}
         \centering
         \includegraphics[width=\textwidth]{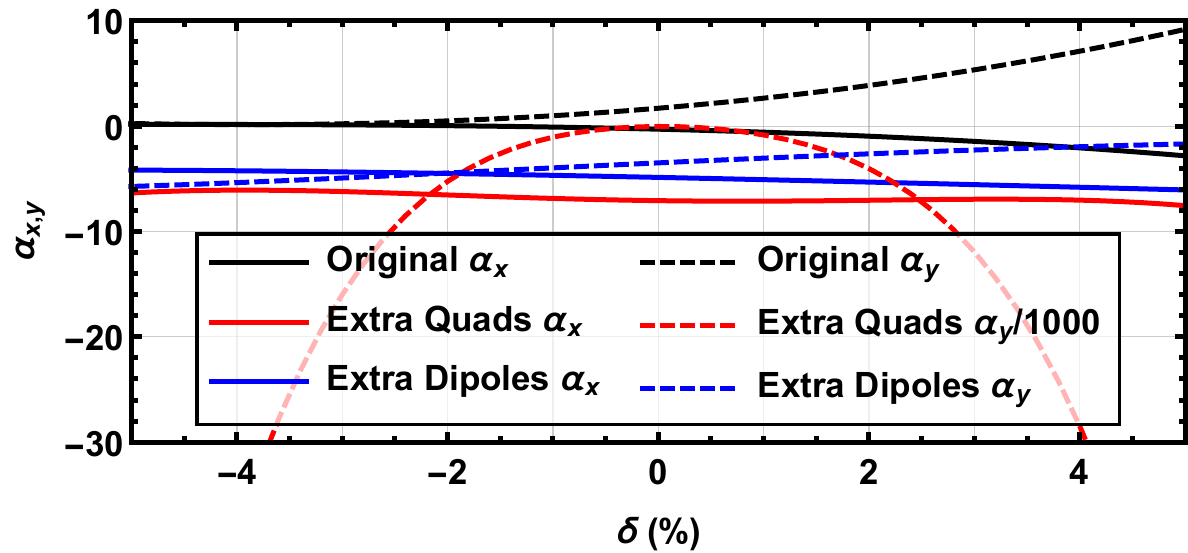}
     \end{subfigure}
        \caption{Chromatic dependence of the beta (top) and alpha (bottom) functions at the end of BC1 for the original, fixed lattice (black), additional quadrupole (red) and additional dipole (blue) solutions.}
        \label{fig:Apochrom}
\end{figure}

In order to study the transverse chromatic properties of the bunches we analyse the chromatic dependences of
the Twiss parameters in Fig.~\ref{fig:Apochrom}, as these determine the behaviour of the chromatic amplitude function. The additional dipole solution shows a smaller chromatic dependence of the Twiss parameters and thus is expected to result in a smaller longitudinal projected emittance growth than the additional quadrupole solution.

\section{Particle Tracking Studies}
\begin{figure}[h]
     \centering
     \begin{subfigure}[h]{0.50\textwidth}
         \centering
         \includegraphics[width=\textwidth]{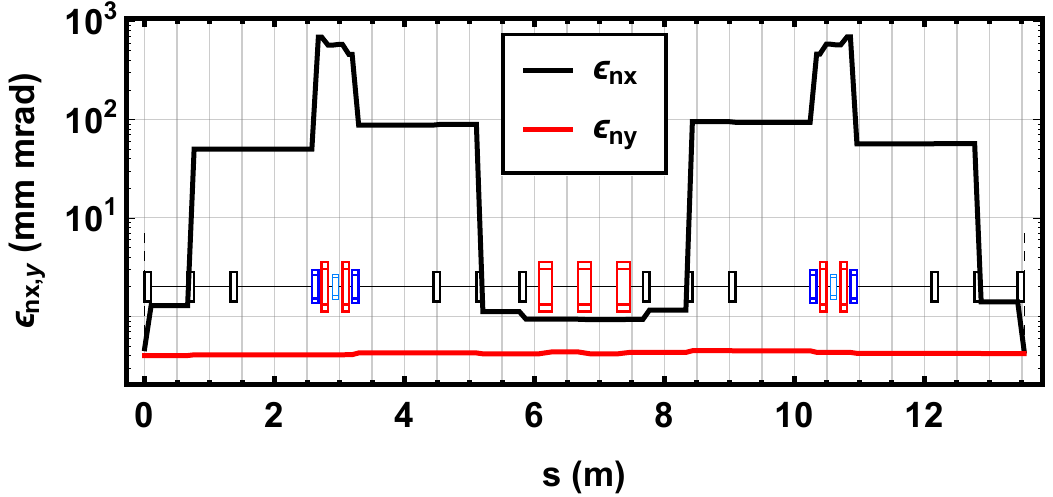}
         \caption{Beam emittance evolution in additional dipole solution.}
         \label{fig:Tracking:Dip}
     \end{subfigure}
     \\
     \begin{subfigure}[h]{0.50\textwidth}
         \centering
         \includegraphics[width=\textwidth]{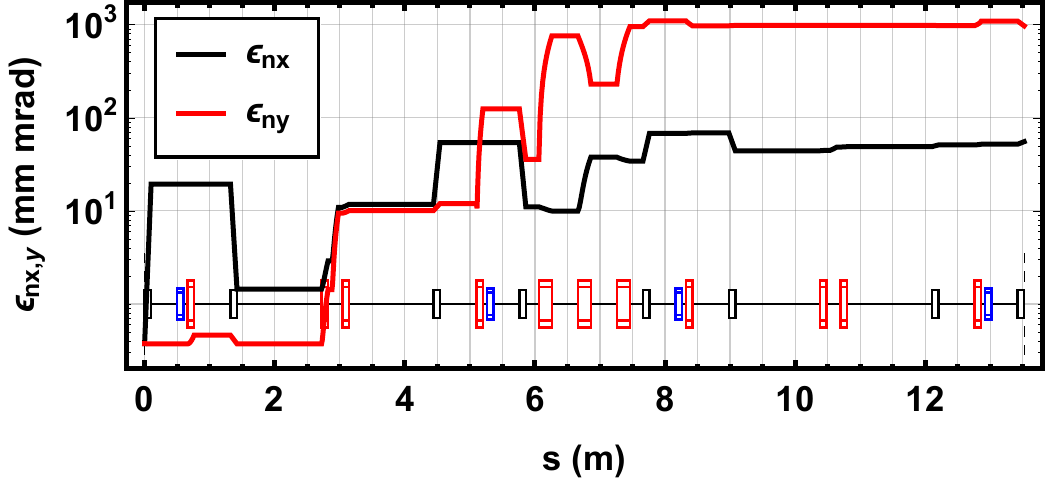}
         \caption{Beam emittance evolution in additional quadrupole solution.}
         \label{fig:Tracking:Quad}
     \end{subfigure}
\caption{Beam emittance evolution in additional dipole and additional quadrupole configurations.}
\label{fig:Tracking}
\end{figure}

In order to further investigate the performance of both candidate lattices we track a bunch from the linac up to the exit of the bunch compressor. We include collective effects, in particular coherent synchrotron radiation (CSR) in this tracking. The effect is however negligible due to the relatively long bunch length.
In the additional quadrupoles solution, as predicted by Fig.~\ref{fig:Apochrom},  emittance growth of orders of magnitude is present. The greatest emittance growth is located in the section before BC1 and especially in the central triplet of the bunch compressor (Fig.~\ref{fig:Tracking:Quad}). Both of these are related to the optical constraints of the system. In order to drive the dispersion and achieve the $R_{56}$ ranges that we are aiming for, all the focusing needed from the quadrupoles results in a strongly focused system. This causes transverse phase space distortion of the large relative energy spread beam, resulting in the emittance growth seen. In addition to this, the beam would be further degraded in the down-stream matching section.
By contrast, the additional bends solution, without the optical constraints, is shown to better preserve the beam quality across the bunch compressor as show in Fig.~\ref{fig:Tracking:Dip}.

\section{Conclusion}
We have presented two solutions to retrofit variable order-by-order momentum compaction into the MAX IV arc-like bunch compressors. In both cases we are able to move from the natural $R_{56}$ through the isochronous condition to the opposite sign (thereby mimicking a chicane-like compressor). In the additional quadrupole case, the strong focusing needed for the dispersion manipulations greatly worsen the beam quality with emittance growths of orders of magnitude. In the additional dipole case, we can do this without chromatic penalty and showing minimal emittance growth using sextupole and octupole magnets. This solution is presently under detailed engineering consideration for adoption in the MAX IV Soft X-ray FEL upgrade project.
\section{Acknowledgements}
The authors wish to thank Simone DiMitri (Elettra Sincrotrone Trieste), Bruno Muratori (STFC Daresbury Laboratory), Pedro Tavares and Erik Mansten (MAX IV) for useful discussions. 

\bibliographystyle{apsrev}
\bibliography{addbend}
\end{document}